\documentclass[trackchanges]{aastex701}
\usepackage{xeCJK}
\usepackage[utf8]{inputenc} 

\usepackage{threeparttable}
\usepackage{adjustbox}
\usepackage{float}
\usepackage{graphicx}
\usepackage{color} 

\begin{document}

\title{Investigation of White-light Emission in Compact Flares}

\author[0000-0002-9961-4357,sname=Song,gname=Yongliang]{Yongliang Song (宋永亮)}
\affiliation{State Key Laboratory of Solar Activity and Space Weather,National Astronomical Observatories, Chinese Academy of Sciences, Beijing 100101, China}
\email[show]{ylsong@nao.cas.cn}


\begin{abstract}

White-light flares (WLFs) are usually tend to be those very large flares. Nevertheless, several small and compact WLFs have been reported and thought to be produced by low-height magnetic reconnection. However, whether low-height magnetic reconnection can efficiently produce WLFs remains unclear. For the first time, we conduct a statistical study of the WL emission in compact flares to address this question. Using over a decade observations from the \textit{Solar Dynamics Observatory} (SDO), we identify 28 compact flares, including 19 C-class and 9 B-class flares. We find these compact flares can be classified into three types based on the magnetic configuration of the flare, corresponding to the U-shape loop (type I), the flux emergence near sunspot (type II), and the fan-spine like structure (type III). For each type, the flares numbers are 9 (7 C-calss and 2 B-class), 9 (3 C-calss and 6 B-calss) and 10 (9 C-calss and 1 B-calss), respectively. We find the occurrence rate of WLFs in compact flares is $\sim60.7\%$ (17/28), and for the C-class the rate can be up to $\sim89.5\%$ (17/19). No WLF was found in B-class compact flares. The occurrence rates for three types are  $\sim77.8\%$ (7/9),  $\sim11.1\%$ (1/9) and 90\% (9/10), respectively. And for the C-class flares, the occurrence rates for three types are 100\% (7/7),  $\sim33.3\%$ (1/3) and 100\% (9/9), respectively. Our results suggest type-I and type-III compact flares are more likely to produce WL emissions. 

\end{abstract}


\section{Introduction} 

White-light flare (WLF), which exhibit enhancement in white-light continuum emission, was first observed in 1859 by Richard C. Carrington \citep{Carrington1859} and Richard Hodgson \citep{Hodgson1859}. WLFs are crucial for understanding the heating processes in the lower solar atmosphere and the mechanisms of stellar flares \citep[e.g.][]{Neidig1989, Ding1999b}. Despite their importance, WLFs are very rare in observations \citep[e.g.][]{McIntosh1972, Neidig1983, Fang2013}. Since the launch of the \textit{Solar Dynamics Observatory} (SDO), we do observe more WLFs comparing with the past\citep[e.g.][]{Song2018b, Jing2024, Cai2024, Li2024b}, but the number is still small. More than 160 years later, some fundamental questions about WLFs still lack definitive conclusions \citep{Hudson2016}. Such as the emission mechanisms and origins of WL enhancement \citep[e.g.][]{Neidig1984, Aboudarham1986, Ding1994, Fang1995}, the energy transportation \citep[e.g.][]{Neidig1989, Ding1999b} and the heating processes for the lower solar atmosphere \citep[e.g.][]{Hudson1972, Aboudarham1989, Machado1989, Metcalf1990, Gan1994, Fletcher2008, Heinzel2014, Kowalski2017}.

Based on the observations, WLFs are usually classified into two types \citep{Fang1995, Ding1999a, Ding1999b}. Type-I WLFs always show strong spatiotemporal correlations between WL enhancements and hard-X ray (HXR) emissions \citep[e.g.][]{Metcalf2003, Chen2005, Chen2006, Hao2012, Cheng2015, Song2018a, Li2024a}. And their flare spectra are often characterized by a Balmer jump with very strong Balmer lines \citep[e.g.][]{Neidig1984, Fang1995}. Type-II WLFs do not correlate with HXR emissions well \citep[e.g.][]{Ryan1983, Ding1994, Sylwester2000} and their flare spectra do not show an evident Balmer jump with strong Balmer lines. In the type-I WLFs, the WL emission likely originates from nonthermal electrons \citep[e.g.][]{Watanabe2010, Kuhar2016}. While the WL emission in type-II WLFs is thought to originate from direct energy release in the lower atmosphere \citep[e.g.][]{Ding1994, Chen2001}.

Observational evidence supporting the explanation of type-II WLFs as low-height energy release events is very scarce. \citet{Song2020} reported a very compact C2.3 WLF, which exhibits a remarkably bright WL kernel. This WLF presents a UV-burst feature associated with a fast magnetic cancelation between two small adjacent pores, significant lower-chromosphere heating and weak HXR emissions. And the 3D magnetic field structure they reconstructed presents a typical U-shape loop. All these suggest this WLF is powered by low-height magnetic reconnection, providing a strong evidence to the hypothesis that type-II WLFs are produced by magnetic reconnection in the lower atmosphere. Similarly, \citet{Jess2008} reported a C2.0 WLF featuring a remarkably bright WL kernel ($\sim300\%$ WL enhancement) within a short duration of $\sim2$ minutes. And the WL kernel has a diameter of $\sim300$ km. Both two compact WLFs, though small in GOES class, exhibit remarkably bright WL emissions.

Are all compact flares prone to producing WL emission, given their tendency for magnetic reconnection in lower atmosphere? This interesting question motives us to conduct the first statistical study of the WL emission in compact flares. We organize the paper as follows. Section \ref{sec:obs} is the observations. We present the results and discussion in Section \ref{sec:R&D}. Summary is given in Section \ref{sec:sum}.

\section{Observations} \label{sec:obs}

The Helioseismic and Magnetic Imager (HMI; \citealp{Scherrer2012}) and the Atmospheric Imaging Assembly (AIA; \citealp{Lemen2012}) on board the \textit{Solar Dynamics Observatory} (SDO) provide high-quality observations of solar full disk from the photosphere to the corona. HMI data contains the continuum intensity maps, Dopplergrams, line-of-sight (LOS) magnetograms, and vector magnetograms using the spectral line of Fe {\scriptsize I} 6173 \AA. The time cadences are 45 s, 45 s, 45 s, and 12 minutes, respectively. And the spatial pixel size is  $\sim0.5^{\prime\prime}$. AIA observes the Sun in seven extreme-ultraviolet (EUV) passbands (94, 131, 171, 193, 211, 304, 335 \AA) and two ultraviolet (UV) passbands (1600, 1700 \AA). The corresponding temporal cadences for EUV and UV passbands are 12 s and 24 s, respectively. The spatial pixel size for AIA is  $\sim0.6^{\prime\prime}$. Soft X-ray (SXR) flux (1–8Å) obtained by the \textit{Geostationary Orbiting Environmental Satellites} (GOES; \citealp{Hanser1996}) provides the important information about the flare class and its temporal evolution. The HXR data from the \textit{Reuven Ramaty High Energy Solar Spectroscopic Imager} (RHESSI; \citealp{Lin2002}) and the Gamma-ray Burst Monitor (GBM; \citealp{Meegan2009}) on board the \textit{Fermi Gamma-Ray Space Telescope} are also employed to study the relationship between the HXR emissions and WL enhancements.

\begin{figure*}[t!]
\centerline{\includegraphics[trim=0.0cm 0.5cm 0.0cm 1.0cm, width=1.05\textwidth]{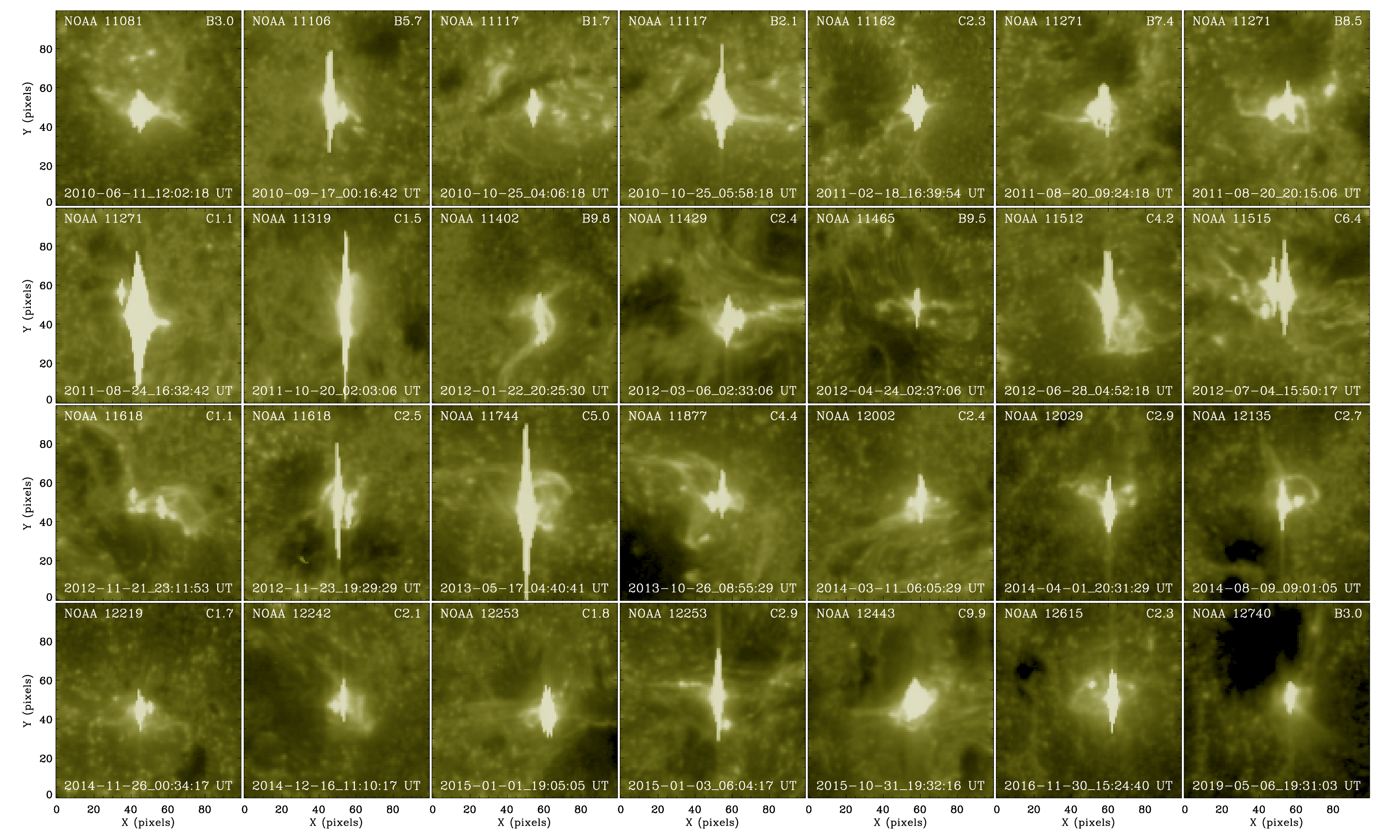}}
\caption{ AIA 1600 \AA\ images at the peak times for 28 compact flares. Detailed information for them is shown in Table \ref{tab1}.
}
\label{fig1}
\end{figure*}

We first obtained the flare list for each year on the GOES website (\url{https://hesperia.gsfc.nasa.gov/goes/goes\_event\_listings/}). Then we downloaded the AIA 1600 \AA\ image at the peak time for each flare. We searched for candidates of compact flares by checking the morphology of each flare at the peak time. After that, we downloaded the AIA 1600 \AA\ movie for each candidate of compact flares. Those exhibit one or multiple dense flare kernels throughout the entire flare process, with no or extremely weak flare ribbons, were ultimately identified as our sample. There are 28 compact flares in our sample including 19 C-class and 9 B-class flares. Figure \ref{fig1} shows these compact flares. It should be noted that the evolution of flare is a complex process. Considering the large number of solar flares, we did not conduct a detailed examination of each flare's dynamic evolution. Therefore, our sample is incomplete. Nevertheless, this limitation does not preclude the validity of our statistical analysis. 

Most compact flares in our sample are very small, with the GOES classes below C5.0. This is consistent with expectations, as compact flares are typically small-scale magnetic reconnection events and therefore release a limited amount of energy. WLFs with the GOES classes below C5.0, are very rare in observations. To date, only a few cases have been reported \citep{Hudson2006, Jess2008, Song2020, Liqiao2024}. WL enhancements, if existing, in these small flares may be too weak to be observed. To improve WL emission detection, we adopted the methodology proposed by \citet{Song2018b} to generate the pseudo-intensity images by magnifying the difference between two adjacent HMI continuum images. Note that this approach is specifically designed to identify the transient enhancements in HMI continuum intensity during the flare. We find 17 of them are WLFs and no WLFs with the B classes.

\section{Results and Discussion} \label{sec:R&D}

\begin{table*}[htbp]
\centering
\caption{List of compact flares}
\label{tab1}
{\tiny
\begin{tabular}{cccccccccccccccccccccc}
\hline
\hline
NOAA & Date & Peak & GOES &  Sunspots & $\Delta T$ & Flare  & \multicolumn{4}{c}{HXR Emission ($counts  s^{-1} cm^{-2} keV^{-1}$)} & HXR & HXR & \multicolumn{2}{c}{WLF} \\
\cline{8-11}
\cline{14-15}
 AR    &     & Time  & Class &   Location    &(min)       & Type & $F_{10 keV}$ & $F_{20 keV}$ & $F_{30 keV}$ & $F_{40 keV}$  & Index($\delta$) & Instr. & (Y/N) &$(dI^{m}_{wl})$ \\
\hline
11081 & 2010.06.11 & 12:02 & B3.0 & N22W41 & 6 & I  & ... & ... & ... & ...&...&...& N& ...\\
11106 & 2010.09.17 & 00:14 & B5.7 & S20W01 & 5 & I  & ... & ... & ... & ...&...&...& N& ...\\
11117	 & 2010.10.25 & 04:06 & B1.7 & N19E12 & 3 & II  & ...& ... &... & ...&...&...& N &  ...\\
11117 & 2010.10.25 & 05:58 & B2.1 & N19E12 & 5 & II & 0.112$\pm$0.008 & 0.001$\pm$0.001 &... &...& 6.5$\pm$0.3 & R & N & ... \\
11162 & 2011.02.18 & 16:39 & C2.3 & N20W04 & 5 & III & ...& ... &... &... &...&...& Y & 0.113\\
11271	 & 2011.08.20 & 09:24 & B7.4	& N16E20 & 5 & II & ...& ... &... &...&...&...& N & ...\\
11271	 & 2011.08.20 & 20:16 & B8.5	& N16E12 & 7& II &1.812$\pm$0.044& 0.016$\pm$0.003& 0.005$\pm$0.002 &...& 9.0$\pm$0.2 & R & N & ...\\
11271	 & 2011.08.24 & 16:33 & C1.1	& N16W36 & 3 & III &1.763$\pm$0.042 & 0.034$\pm$0.004 & 0.007$\pm$0.002 &0.003$\pm$0.002& 6.4$\pm$0.1 & R & Y & 0.092\\
11319	 & 2011.10.20 & 02:03 & C1.5 & N07W59 & 5 & III & 1.446$\pm$0.038 & 0.139$\pm$0.014 & 0.027$\pm$0.011 & 0.014$\pm$0.007 & 4.8$\pm$0.4 & F  & Y &0.070\\
11402 & 2012.01.22 & 20:26 & B9.8 & N33W16 & 5 & II &  ...& ... &... &...&...&...& N & ...\\
11429 & 2012.03.06 & 02:33 & C2.4 & N16E41 & 11 & II & 2.100$\pm$0.051 & 0.131$\pm$0.013 & 0.027$\pm$0.010 & 0.013$\pm$0.006 & 5.1$\pm$0.5 & F & N & ...\\
11465 & 2012.04.24 & 02:37 & B9.5	 & S16W01 & 10 & III & 0.529$\pm$0.016 & 0.019$\pm$0.002 & 0.002$\pm$0.001 &...& 6.1$\pm$0.1 & R & N & ...\\
11512	 & 2012.06.28 & 04:52 & C4.2	 & S16E09 & 10 & I & ...& ... &...&...&...&...& Y & 0.101\\
11515	 & 2012.07.04 & 15:50 & C6.4 & S17W19 & 5 & III & 21.077$\pm$0.434 & 0.743$\pm$0.028 & 0.080$\pm$0.014 & 0.027$\pm$0.008 & 6.5$\pm$0.6 & F & Y & 0.119\\
11618	 & 2012.11.21 & 23:12 & C1.1 & N06W01 &5 & II & 1.714$\pm$0.045 & 0.919$\pm$0.029 & 0.441$\pm$0.019 & 0.131$\pm$0.010 & 3.8$\pm$0.03 & F & N & ...\\
11618	 & 2012.11.23 & 19:30 & C2.5 & N07W27 & 5 & III& 0.110$\pm$0.180 & 3.868$\pm$0.105 & 1.398$\pm$0.042 & 0.380$\pm$0.016 & 4.9$\pm$0.2 & F & Y & 0.101\\
11744 & 2013.05.17 & 04:41 & C5.0 & N06W29 & 6 & I & 1.192$\pm$0.029 & 0.048$\pm$0.003 & 0.007$\pm$0.001 &...& 6.0$\pm$0.4 & R & Y & 0.139\\
11877 & 2013.10.26 & 08:55 & C4.4 & S11W22 & 5 & III & 3.406$\pm$0.092 & 0.038$\pm$0.004 & 0.004$\pm$0.003 &...& 9.1$\pm$0.2 & R & Y & 0.113\\
12002 & 2014.03.11 & 06:05 & C2.4 & S19E34 &6 & I & 1.746$\pm$0.040 & 0.013$\pm$0.002 & 0.002$\pm$0.001 &...& 9.5$\pm$0.7 & R  & Y & 0.159\\
12029 & 2014.04.01 & 20:32 & C2.9	 & N18E20 &6 & I & 1.731$\pm$0.039 & 0.023$\pm$0.003 & 0.004$\pm$0.002 &...& 8.5$\pm$0.3 & R & Y &0.105\\
12135 & 2014.08.09 & 09:01 & C2.7 & N12E34 & 8 & III & ...& ... &...&....&... &...& Y & 0.091\\
12219 & 2014.11.26 & 00:34 & C1.7 & N05E06 & 6 & I & ...&...&.... & ... &...  &... & Y & 0.113\\
12242 & 2014.12.16 & 11:09 & C2.1 & S20E17 & 5 & II & 4.933$\pm$0.110 & 0.073$\pm$0.005 & 0.003$\pm$0.001 &... & 9.6$\pm$0.1 & R & Y & 0.116\\
12253 & 2015.01.01 & 19:05 & 	C1.8 & S05E38 & 4 & III & ...& ... &... &... &...&...& Y & 0.095\\
12253 & 2015.01.03 & 06:04 & C2.9	 & S05E19 & 5 & I & 14.887$\pm$0.308 & 0.445$\pm$0.020 &0.034$\pm$0.010 &0.017$\pm$0.006 & 7.5$\pm$0.5 & F & Y & 0.258\\
12443 & 2015.10.31 & 19:33 & C9.9 & N07E53 & 13 & III & ...& ... &... &... &... &... & Y & 0.202\\
12615 & 2016.11.30 & 15:25 & 	C2.3	 & S07E42 & 5 & I & 19.356$\pm$0.398 & 0.459$\pm$0.019 & 0.072$\pm$0.011 & 0.011$\pm$0.006 & 6.2$\pm$0.4 & F & Y &0.195\\
12740 & 2019.05.06 & 19:30 & B3.0 & N08E43 & 5 & II & ...& ... &... &...&...&...& N & ...\\
\hline
\end{tabular}
\tablecomments{\scriptsize $\Delta T$ refers to flare duration given by GOES 1-8 \AA\ flux. Flare types are classifications of compact flares based on their magnetic topologies, see Figure \ref{fig5}  and the corresponding description in the text for details. $F_{10 keV}$, $F_{20 keV}$, $F_{30 keV}$, $F_{40 keV}$ refer to the HXR flux at 10, 20, 30 and 40 keV. HXR Index($\delta$) is the electron spectral index. HXR Instr. refer to the instruments providing the corresponding HXR observations. `R' represents Rhessi and `F' represents Fermi. For the WLF, `Y' means Yes and `N' means No, repectively. $dI_{wl}^m$ refer to the maximum value of WL enhancement calculated by $(I_p - I_0)/I_0$. Here $I_p$ and $I_0$ refer to the HMI continuum intensities when the WL enhancement reaches the peak and before its appearance, respectively.
}
}
\end{table*}


\begin{figure*}[!htbp]
\centerline{\includegraphics[trim=0.0cm 0.5cm 0.0cm 0.0cm, width=0.96\textwidth]{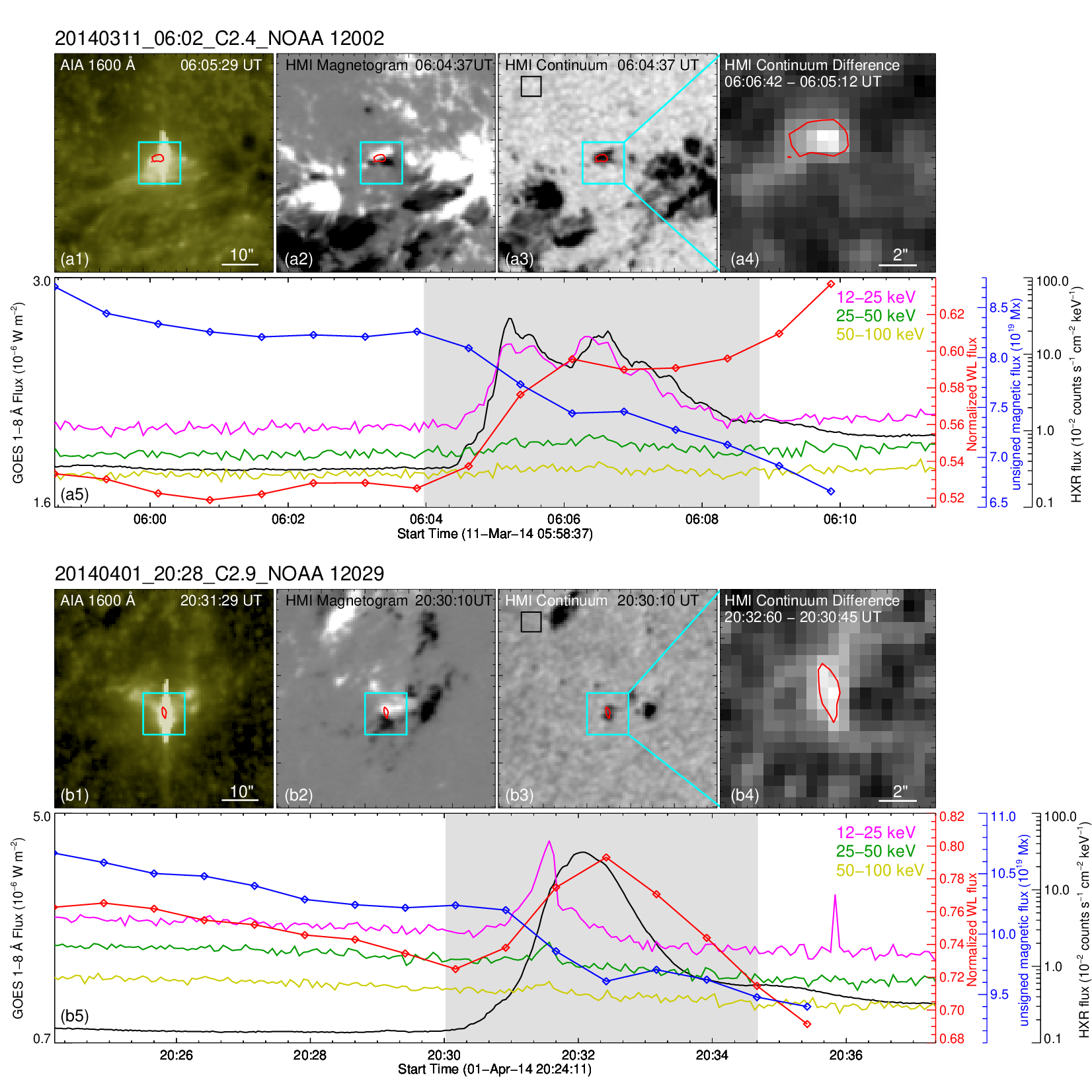}}
\caption{Two samples for type-I compact flares. The above is for the C2.4 flare occurred in NOAA active region (AR) 12002. The bottom is for the C2.9 flare occurred in NOAA AR 12029. (a1)-(a4) and (b1)-(b4) show the AIA 1600 \AA\ images around the flare peak, the HMI line-of-sight magnetograms, HMI continuum images and continuum difference images ($(I_p - I_0)/I_0$), respectively. The cyan boxes in these panels mark the flare regions, where we calculate the unsigned magnetic fluxes. The red contours show the white-light (WL) enhancement areas with the level of $(I_p - I_0)/I_0 > 0.08$. The black boxes in panels (a3) and (b3) mark the quiet regions which used to normalize the WL flux ($I_{wl}/I_q$). (a5) and (b5) show the temporal variations of the normalized WL fluxes (red), unsigned magnetic fluxes (blue), GOES 1-8 \AA\ fluxes (black) and Hard X-ray (HXR) emissions (magenta, green and yellow). The gray shadings indicate the flaring periods.
}
\label{fig2}
\end{figure*}

\begin{figure*}[!htbp]
\centerline{\includegraphics[trim=0.0cm 0.5cm 0.0cm 0.0cm, width=0.96\textwidth]{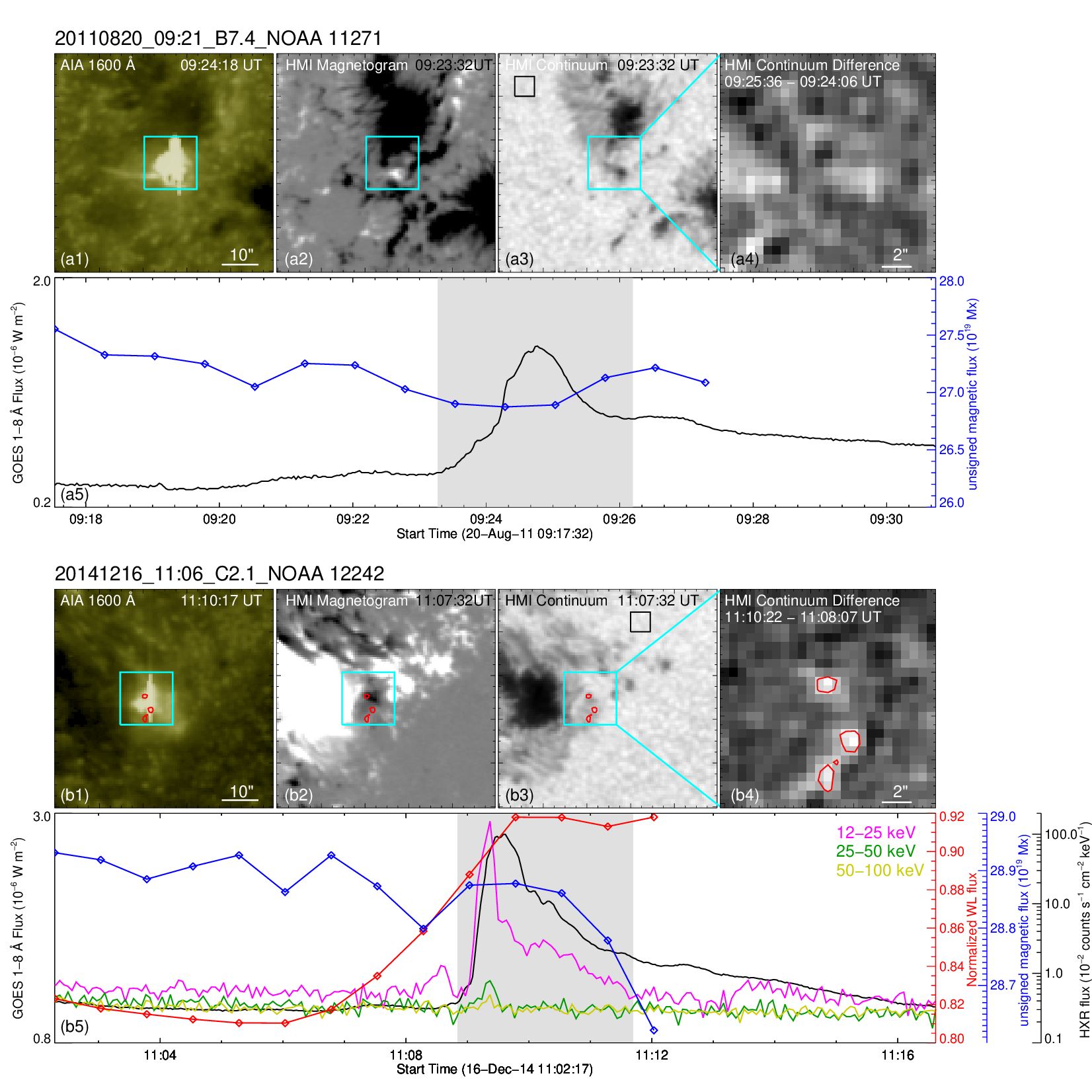}}
\caption{Similar to Figure \ref{fig2}, but two samples for type-II compact flares.
}
\label{fig3}
\end{figure*}

\begin{figure*}[!htbp]
\centerline{\includegraphics[trim=0.0cm 0.5cm 0.0cm 0.0cm, width=0.96\textwidth]{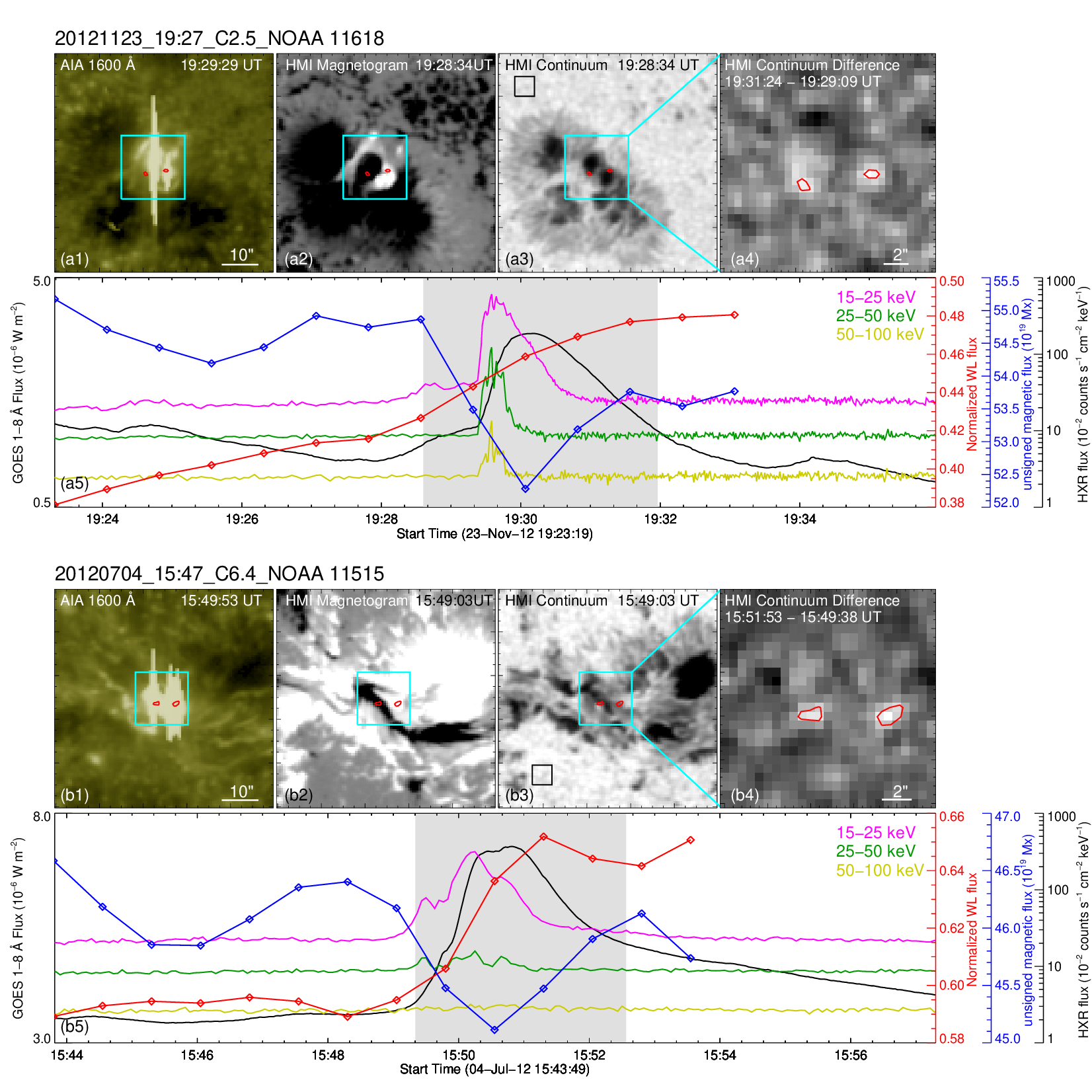}}
\caption{Similar to Figure \ref{fig2}, but two samples for type-III compact flares.
}
\label{fig4}
\end{figure*}

\begin{figure*}[!htbp]
\centerline{\includegraphics[trim=0.0cm 1.5cm 0.0cm 1.5cm, width=0.98\textwidth]{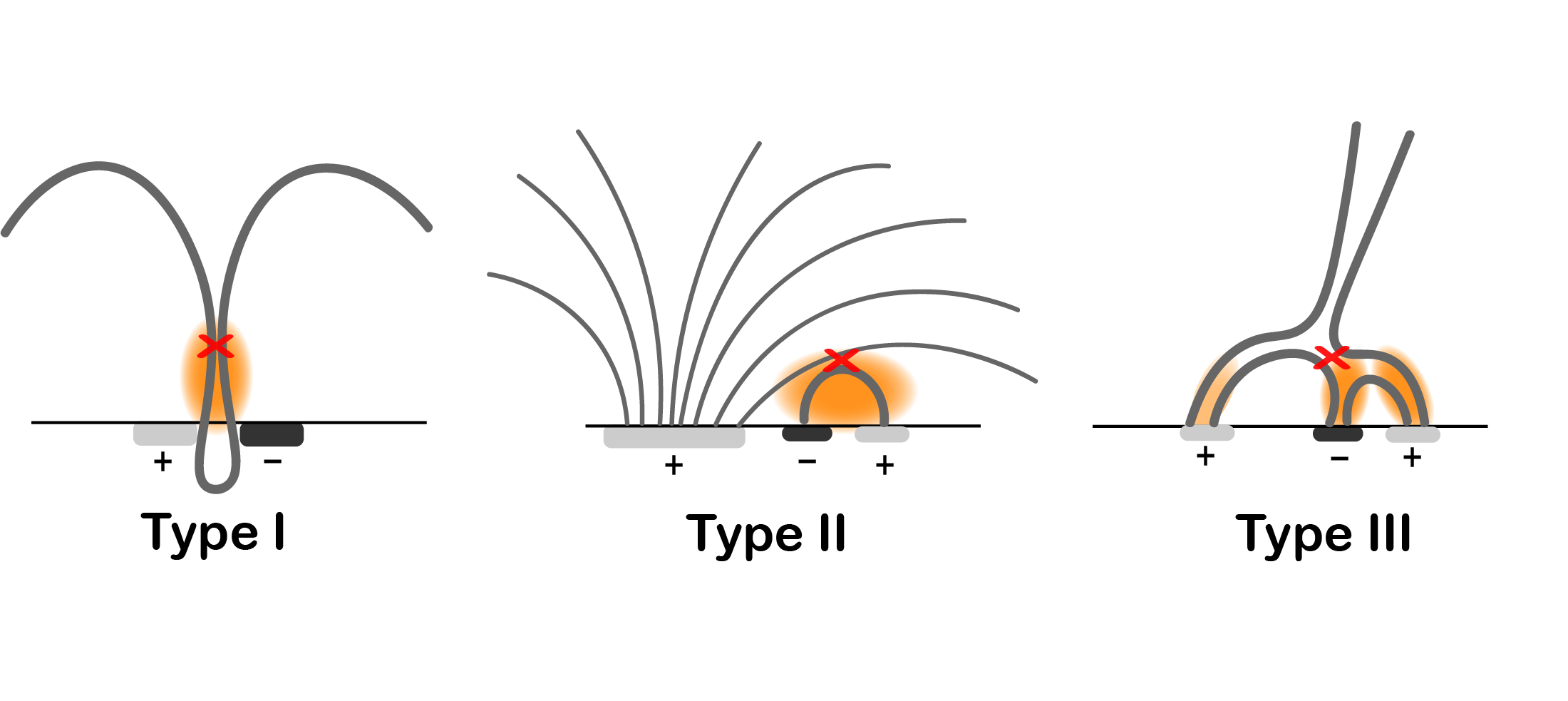}}
\caption{Catoon shows the three types of compact flares. ``X" refer to the magnetic reconnection. While the yellow areas refer to the flare emissions. Grey lines represent the magnetic flux tubes.
}
\label{fig5}
\end{figure*}

Figure \ref{fig1} shows the AIA 1600 \AA\ images at the peak times for the 28 compact flares. We see all these flares displaying very bright flare kernels. It is hard to see any feature of flare ribbons in them. It can be also found that some of these compact flares exhibit two or more bright kernels instead of just one, such as the C6.4 in active region NOAA 11515, the C2.5 in NOAA 11618, the C4.4 in NOAA 11877 and the C2.7 in NOAA 12135. And some of these compact flares can be clearly seen occurred near a big sunspot, such as the B3.0 in NOAA 12740, the B7.4 and B8.5 in NOAA 11271. Table \ref{tab1}  presents the detailed information of all these compact flares. Interestingly, we find all these compact flares can be classified into three types. It should be noted that the classification here is for compact flares, not for WLFs.

Figure \ref{fig2} shows two cases for the type-I compact flares, i.e., the C2.4 in NOAA 12002 and the C2.9 in NOAA 12029. Both two flares show only one bright kernel from AIA 1600 \AA\ image. It is clearly seen these two flares occurred between two small pores with different polarities of magnetic field (see panels (a2)-(a3) and (b2)-(b3)). Both two flares are WLFs, and we see the WL enhancement also presents a bright kernel just located at the magnetic polarity inversion line (PIL). Though merely C2-class flares, the two flares exhibited significant WL enhancements reaching $\sim15.9\%$ and $\sim10.5\%$, respectively.

Panels (a5) and (b5) in figure \ref{fig2} show the temporal evolutions of WL emission (red), unsigned magnetic flux in the flaring region (blue), GOES 1-8 \AA\ SXR flux (black) and HXR fluxes (magenta, green and yellow) for the flares. Grey shadings indicate flare durations. We see obviously magnetic cancellations during the flares. For the C2.4 flare in NOAA 12002, the unsigned magnetic flux decreased about 13.7\% from $\sim8.26\times10^{19}$ Mx to $\sim7.13\times10^{19}$ Mx. The magnetic cancellation rate is $\sim4.19\times10^{16}$ Mx/s. For the C2.9 flare in NOAA 12029, the unsigned magnetic flux decreased about 7.4\% from $\sim10.24\times10^{19}$ Mx to $\sim9.48\times10^{19}$ Mx. And the corresponding magnetic cancellation rate is about $2.41\times10^{16}$ Mx/s. HXR emissions in both two events are very weak. For the C2.4 flare in NOAA 12002, the HXR emission above 25 keV is extremely faint, becoming virtually indistinguishable from the background. And for the C2.9 flare in NOAA 12029,  we can see very weak transient impulse in HXR 25-50 keV flux. The normalized WL flux presents a good correlation with the temporal HXR and SXR emissions for the C2.4 flare in NOAA 12002. However, there is no such temporal correlations in the the C2.9 flare of NOAA 12029. The peak of normalized WL flux lags the peaks of HXR and SXR fluxes about 1 and 0.5 minutes, respectively.

Thus, we can conclude that the type-I compact flare usually occurs between two small pores with different polarities of field, associated with fast magnetic cancellation. It often displays as one very bright and compact flare kernel above the PIL of two pores. The WL enhancement in such flare also presents as one bright kernel located above the PIL. The HXR emission is usually very weak. It may have or no temporal correlations with the WL flux. All the above features coincide with the C2.3 WLF reported by \citet{Song2020}. They analyzed the magnetic field structure of the flare, giving a very typical U-shape loop above two small pores. Therefore, we propose that type-I compact flares share a U-shape loop structure (see Figure \ref{fig5}). 

Figure \ref{fig3} shows two samples for the type-II compact flares. The B7.4 flare in NOAA 11271 is a normal flare without any detectable WL enhancement. While the C2.1 flare of the NOAA 12242 is a WLF. Both two flares occurred near a big sunspot. Unlike the type-I compact flares, the flare kernels in the two flares seem not very intensely compact. We see no evident magnetic cancellations during the flare in both two cases. Unlike type-I compact WLFs, three bright kernels for the WL enhancement in the C2.1 flare of NOAA 12242 can be seen (see panels (b1)-(b4)). One locates in the negative magnetic field and two locate in the positive magnetic field near a big sunspot. The max value for the WL enhancement in this C2 WLF is  $\sim11.6\%$.

Panel (b5) shows temporal variations of the WL emission, unsigned magnetic flux, SXR and HXR fluxes for the  C2.1 flare in NOAA 12242. Similarly, we see the HXR exhibit significant radiation in the 12–25 keV range, while showing a weak response in the 25–50 keV range. Above 50 keV, the HXR radiation is nearly negligible compared to the background. And the peak of WL flux also lags the HXR and SXR peaks about 30 s and 10 s, respectively. No HXR emissions recored available for the B7.4 compact flare in NOAA 11271 whether by RHESSI or FERMI. 

Thus, we see the type-II compact flares usually occur near sunspot. Some of them can be WLFs. And the WL enhancement may present as several kernels located in positive and negative magnetic fields, respectively. The HXR emissions in this type compact flares are also very weak. And there is no strong correlations between the HXR emissions and WL flux in this type of compact flares. This type of compact flare should be resulted from the reconnections between emerging flux and the above magnetic field from the penumbra of nearby sunspot, like \citet{Liusuo2014} shows.

Figure \ref{fig4} shows two type-III compact flares, i.e., the C2.5 flare in NOAA 11618 and the C6.4 in NOAA 11515. Both two flares are WLFs. Unlike previous two types of compact flares, we can see two  bright and intense kernels from AIA 1600 \AA\ images in this type of compact flares. The WL enhancements in both two flares also present as two bright kernels. The max values for the WL enhancements in two WLFs are $\sim10.1\%$ and $\sim11.9\%$, respectively. The photosphere magnetic fields in both cases share a similar feature, that one polarity magnetic field surrounded by another polarity of magnetic field, in a shape of circle or ribbons. Two flare kernels and corresponding WL enhancements just located in two polarities of magnetic field. This type of magnetic field structure called fan-spine shape like structure, which often result in circular-ribbon flares \citep[e.g.][]{Song2018c, Zhangqm2024} or triple-ribbon flares \citep[e.g.][]{Wanghaimin2014}.

Panels (a5) and (b5) give the temporal variations of the WL emission, unsigned magnetic flux, SXR and HXR fluxes for two flares. We can see very obviously magnetic transients during the flare in both two cases, which are commonly thought to be artifacts due to the distortions of the observational line profile \citep[e.g.][]{Qiu2003, Song2018b}. The unsigned magnetic flux in the C2.5 of NOAA 11618 slightly decreased about 2\% from $\sim5.59\times10^{20}$ Mx to $\sim5.38\times10^{20}$ Mx. While there is no obviously magnetic flux decrease for the C6.4 flare in NOAA 11515. HXR emissions in the C2.5 of NOAA 11618 are obviously stronger than that in the C6.4 flare in NOAA 11515. We see an obviously emission in 50-100 keV flux in the former flare, while weak emission in the 25-50 keV flux and no detectable emission in 50-100 keV flux for the latter flare. There are also no good temporal correlations between the HXR fluxes and the WL enhancements in both cases. The peak of WL enhancement of the C6.4 flare in NOAA 11515 lags the HXR and SXR fluxes about 1 min and 30 s, respectively. For the C2.5 flare in NOAA 11618, the WL emission can be seen superimposed on a continuously brightening background, making it relatively difficult to determine the peak time of the WL emission. However, it is evident that the peak time of the WL emission still significantly lags behind the peaks of the HXR and SXR fluxes in this flare.

For the type-III compact flares, we see they usually occur associated with fan-spine like magnetic field, showing two bright flaring kernels. The WL enhancements also present as two bright kernels located in two different polarities of magnetic field. HXR emission in this type of flare is generally quite weak, some of them may exhibit a radiation in the 50–100 keV range. \citet{Chitta2017} has been reported a compact solar UV burst occurred in a magnetic field with a fan-spine topology. They calculated the height of the 3D magnetic null point that about 500 km above the photosphere. And the EUV emission at the coronal loops' footpoints presents as a UV-burst.

Figure \ref{fig5} schematically illustrates the magnetic field configurations for three types of compact flares. For the type-I compact flares, magnetic reconnection occurs along the inner side of the U-shape loop. Energy transports downward and heats the lower atmosphere in a concentrated region above the PIL between two small pores. That is why we can see just one bright kernel of WL enhancement in this type compact WLF (Figure \ref{fig2}). For the type-II compact flares, small-scale magnetic flux emerges near a sunspot and reconnects with magnetic field of the penumbra of the sunspot. The released energy transport downward mainly along two legs of the small-scale loop. Then we can see WL enhancements in both two polarities of the small-scale loop rather than just one kernel in this type compact WLF (Figure \ref{fig3}). For the type-III compact flares, magnetic reconnection occurs at null point. Energy transport downward along the inner spine and surrounding loops. The spine structure may be not symmetric, therefore the heating in surrounding loops is non-uniform. That is why, in such events, we can observe more than one bright compact kernels and other weak bright structures (Figure \ref{fig4}).

\begin{table*}[!htbp]
\centering
\caption{Occurrence rate of WLFs for compact flares}
\label{tab2}
{
\begin{tabular}{cccccc}
\hline
\hline
Flare Type & Flares & WLFs &  Occurrence rate & Occurrence rate for C-class flares\\
\hline
I  & 7C+2B  & 7C  & 77.8\%  & 100\% \\
II & 3C+6B & 1C   & 11.1\% & 33.3\% \\
III & 9C+1B & 9C  & 90.0\% &100\% \\ 
\hline
Total & 19C+9B & 17C  & 60.7\% & 89.5\%\\
\hline
\end{tabular}
}
\end{table*}

Table \ref{tab2} gives the occurrence rate of WLFs in compact flares. We see the probability of WLFs occurring in compact flares is  quite high with a rate $\sim60.7\%$. And for the C-class compact flares, the occurrence rate can reach to $\sim89.5\%$. No B-class WLF was found in our samples, suggesting that sufficient energy is required to produce WLF even for compact flares. In our sample, 7 C-class and 2 B-class,  3 C-class and 6 B-class, and 9 C-class and 1 B-class compact flares are classified as type-I, type-II, type-III compact flares, respectively. Type-I and type-III compact flares seams more likely producing WLFs, with a rate of $\sim77.8\%$ and 90.0\%, respectively. And the occurrence rates for the C-class flares of both types are 100\%. Only one C-class WLF appeared among Type-II compact flares. The occurrence rate is just $\sim11.1\%$ and for the C-class is $\sim33.3\%$.

\begin{figure*}[ht!]
\centerline{\includegraphics[trim=0.0cm 0.5cm 0.0cm 0.0cm, width=0.96\textwidth]{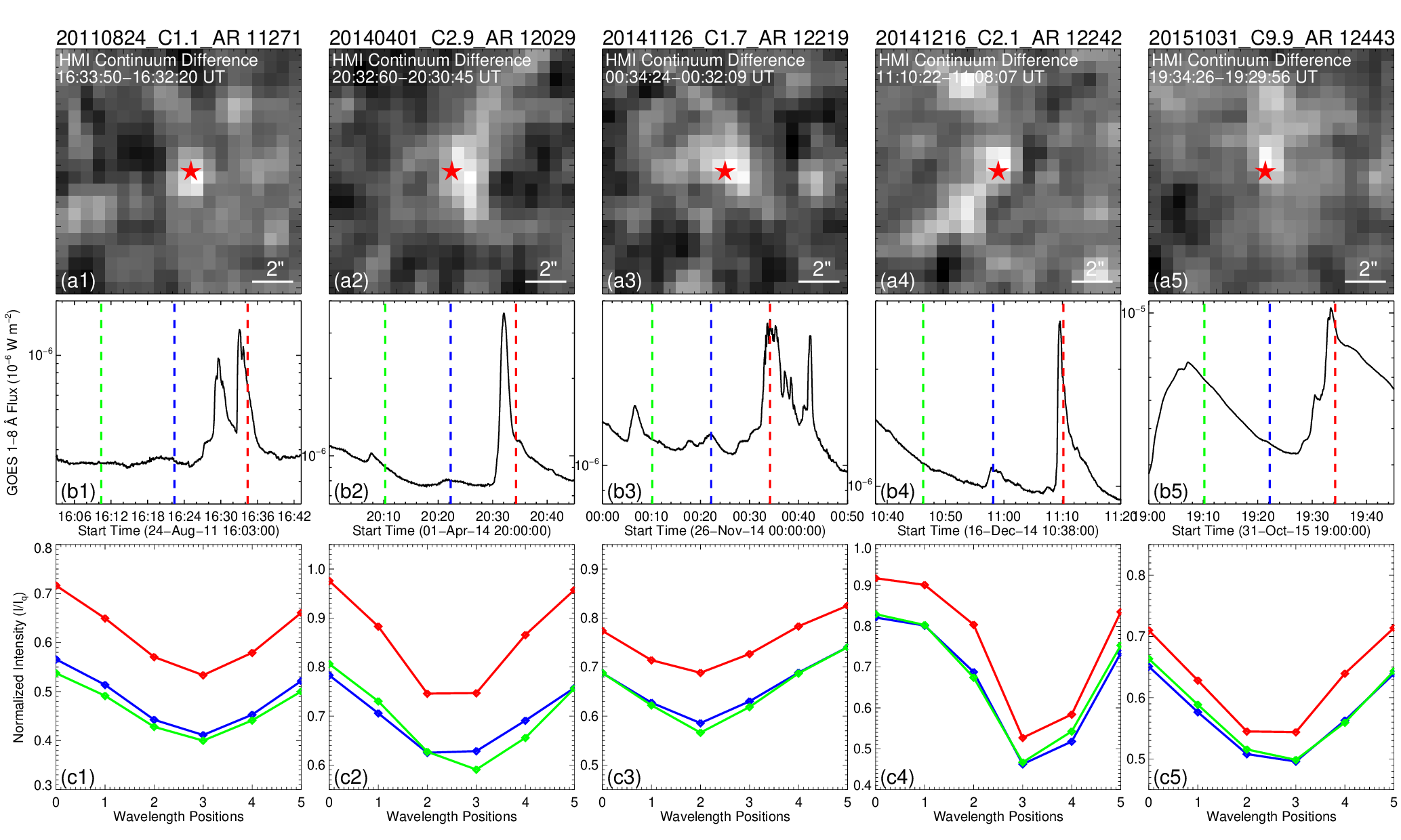}}
\caption{Spectral profiles of HMI Fe {\scriptsize I}  6173 \AA\ at the center of WL kernels for five flares. (a1)-(a5) show the WL kernels of these flares. The red stars mark the positions where we show the HMI Fe {\scriptsize I}  6173 \AA\ spectral profiles. (b1)-(b5) give the GOES 1-8 \AA\ fluxes of these flares. Vertical dashed lines marks the times for these spectral profiles. Red is during the flare, blue and green are before the flare. (c1)-(c5) show these corresponding HMI Fe {\scriptsize I}  6173 \AA\ spectral profiles. All these spectral profiles are normalized by using the intensities in quiet regions.
}
\label{fig6}
\end{figure*}

\begin{figure*}[ht!]
\centerline{\includegraphics[trim=0.0cm 0.5cm 0.0cm 0.0cm, width=0.96\textwidth]{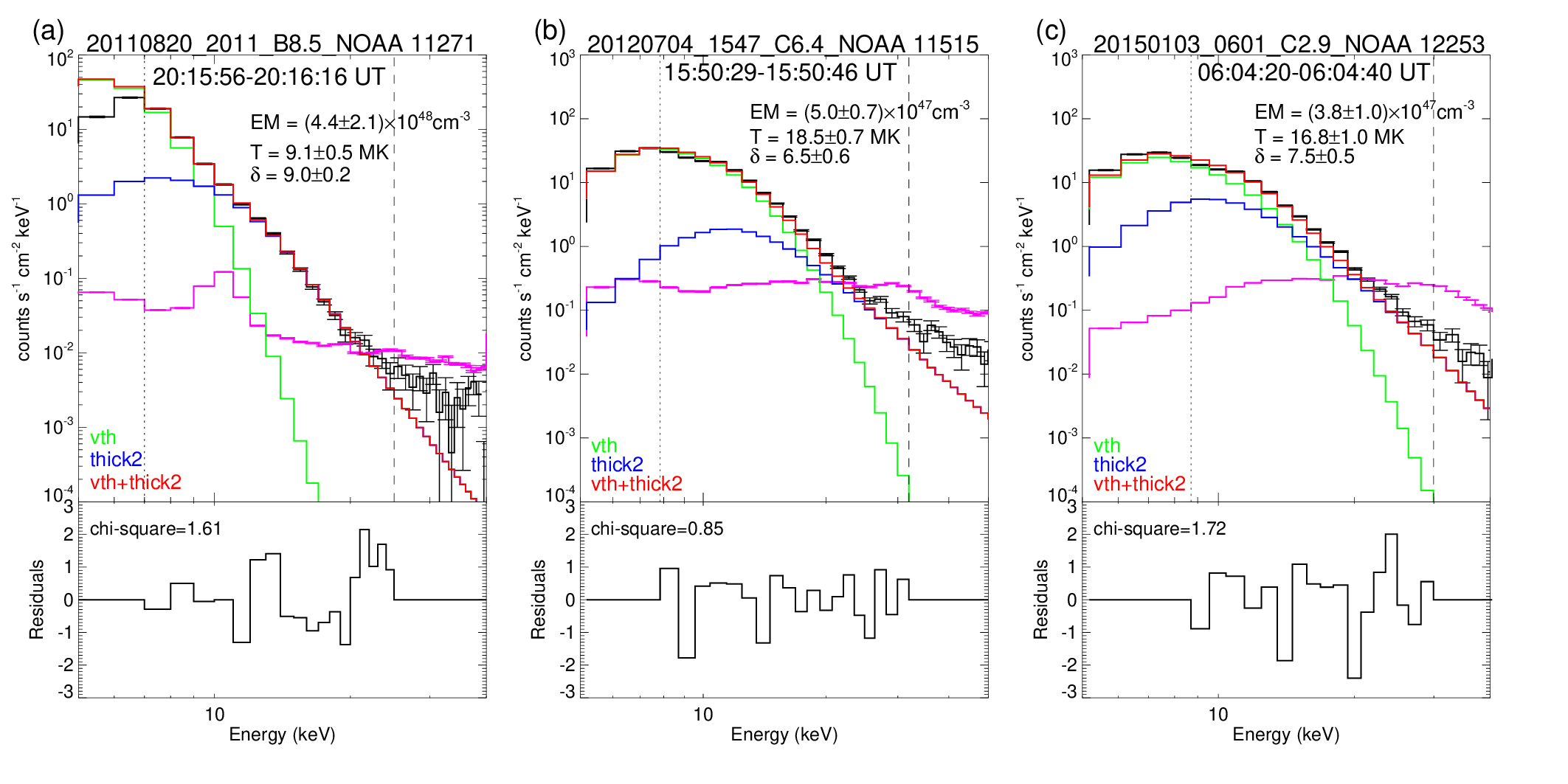}}
\caption{HXR spectral fittings using a thermal (vth, green) and nonthermal model (thick2, blue) for three flares. A single power-law model was used in the spectral fitting. The black curves are the observational count spectrums after subtracting the background (magenta curves).The time interval and fitting parameters are shown in each panel. The fitting energy ranges are indicated by the two vertical lines (dotted and dashed). The spectrum fitting residuals are also given in each panel.
}
\label{fig7}
\end{figure*}

\begin{figure*}[ht!]
\centerline{\includegraphics[trim=0.0cm 0.5cm 0.0cm 0.0cm, width=0.96\textwidth]{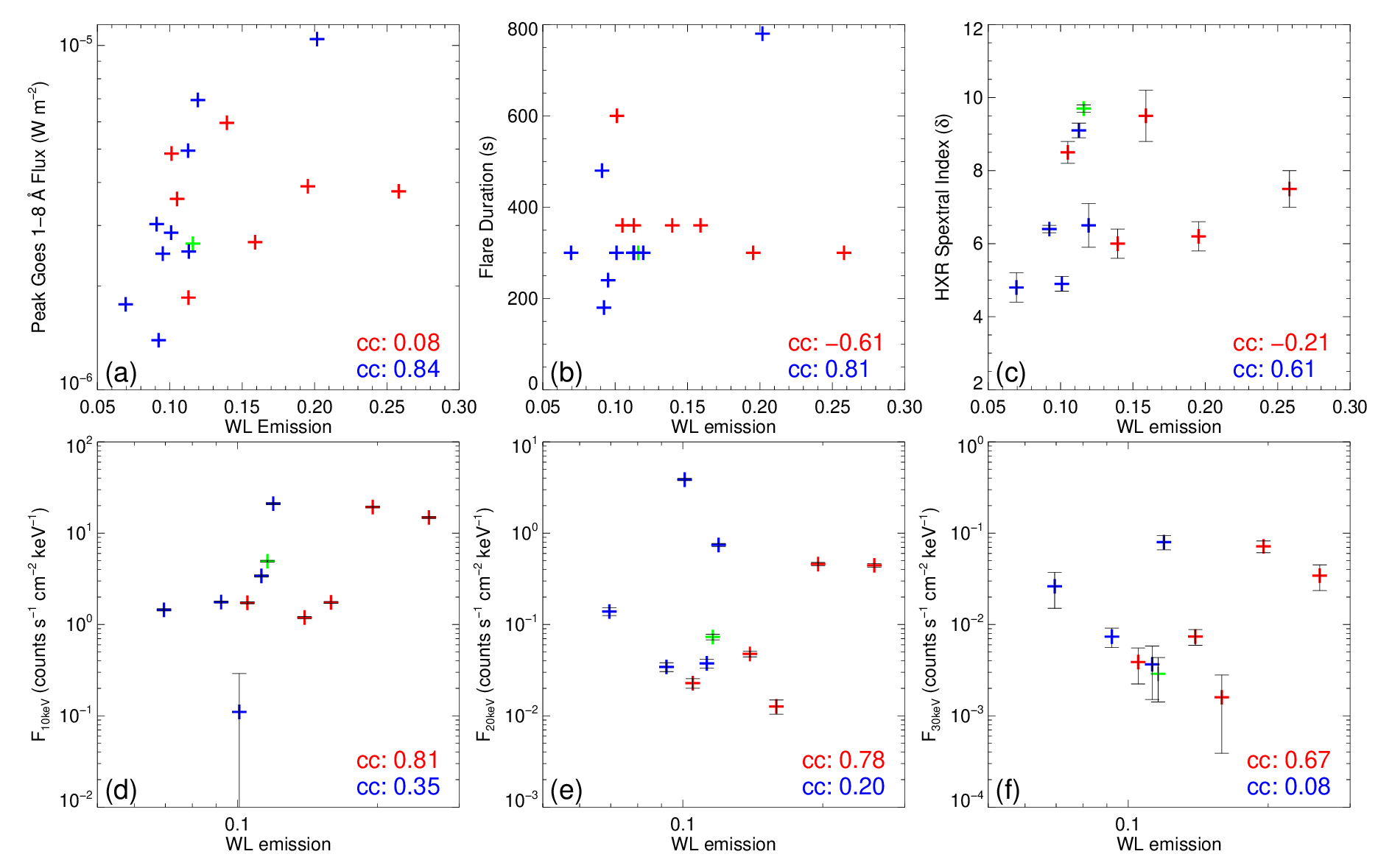}}
\caption{Scatter plots the WL emissions ($(I_p - I_0)/I_0$) and the peak GOES 1-8 \AA\ fluxes (a), flare durations (b), HXR spectral indexes (c), the HXR fluxes at 10, 20 and 30 keV, i.e., $F_{10keV}$(d), $F_{20keV}$(e) and $F_{30keV}$(f). Red, green and blue ``+'' in each panel represent the type-I, type-II and type-III compact WLFs, respectively. The corresponding correlation coefficients for type-I (red) and type-III (blue) compact WLFs are given in each panel.
}
\label{fig8}
\end{figure*}

In recent years, HMI data have played a crucial role in the statistical studies of WLFs \citep[e.g.][]{Kuhar2016, Song2018c, Castellanos2020, Jing2024, Cai2024, Li2024b}. However, it is important to note that the HMI continuum is constructed from six spectral sampling points, using a specific formula as following:
\begin{equation}
\emph{$I_{c}=\displaystyle\frac{1}{6}~\displaystyle{\sum_{j=0}^5}~[I_j+I_d~exp(-\frac{(\lambda-\lambda_0)^2}{\sigma^2})]$}\label{equation1}
\end{equation}
here $\lambda_0$, $\sigma$ and $I_d$ are the rest wavelength, line width and line depth, estimated by taking six sampling points ($I_j$) across the Fe {\footnotesize I} line of 6173 \AA~ \citep{Couvidat2012}.

\citet{Svanda2018} compared the observed Fe {\scriptsize I}  6173 \AA\ spectral line profiles from HMI with profiles of the same line synthesized from model atmospheres. These model atmospheres were retrieved by applying an inversion code to the Stokes profiles observed by the Solar Optical Telescope \citep[SOT;][]{Tsuneta2008} on board \emph{Hinode} \citep{Kosugi2007}. They found that the HMI continuum intensities agree well with those from \emph{Hinode} in quiet-Sun regions. However, in regions with stronger magnetic fields, the spectral lines become highly complex and can even change into emission lines during flares. In such cases, the six sampling points used by HMI are insufficient to accurately represent the true line profile, leading to significant deviations between the constructed and the actual continuum intensities. It is worth noting that their study focused on a major X9.3‑class flare. For M‑class or smaller C‑class flares, the behavior of HMI spectral lines—specifically the extent of their deviation from the true profiles—still requires further investigation, particularly through comparisons with more accurate continuum observations.

It is important to note that \citet{Hong2018} used RADYN simulations \citep{Carlsson1992, Carlsson1995, Carlsson1997, Carlsson2002, Allred2005, Allred2015} to demonstrate that when the photosphere and lower chromosphere are effectively heated during flare, the HMI Fe {\scriptsize I} 6173 \AA\ six-point spectral lines show a clear enhancement in radiation. More recently, \citet{Granovsky2025} also found that when the HMI 6173 Å line changes from an absorption to an emission profile, the corresponding cool photospheric layers are effectively heated. These results indicate that when the six-point HMI 6173 \AA\ radiation is significantly enhanced during a flare, it implies effective heating of the lower atmosphere. Given that the generation of WLF is primarily associated with such heating of the lower solar atmosphere, an evident emission enhancement in the HMI 6173 \AA\ spectral line during a flare can thus serve as an indicator of WLF occurrence.

Figure \ref{fig6} displays the HMI Fe {\scriptsize I} 6173 \AA\ spectral profiles within the WL core regions for five flares.  To date, only a limited amount of high-cadence HMI Stokes and magnetic field data (hmi.S\_90s and hmi.B\_90s) is publicly available (see \url{http://jsoc.stanford.edu/data/hmi/highcad/}). Among the events in our sample, only the C2.3 flare in NOAA Active Region (AR) 12615 has corresponding high-cadence data. \citet{Song2020} has already analyzed this event and demonstrated an evident radiative enhancement across all six HMI sampling points in the flare kernel. So the spectral profiles presented in Figure \ref{fig6} are all from the standard 12-minute cadence Stokes data. Since compact flares typically have short durations ($<12$ minutes), only five flares in our sample have corresponding Stokes measurements during flare. All such cases are included in Figure \ref{fig6}.

It can be clearly observed that the radiation intensities at the six HMI spectral line positions remain similar or exhibit only minor variations before flare. In contrast, during the flare, the spectral radiation is substantially enhanced at every sampling point, consistent with the simulation results of \citet{Hong2018} and \citet{Granovsky2025}. This indicates that all these flares are accompanied by significant heating of the lower solar atmosphere --- a process closely linked to WLF generation.

Figure \ref{fig7} presents the spectral fitting results for three flares: the B8.5 flare from NOAA AR 11271, the C6.4 flare from NOAA AR 11515 and the C2.9 flare from NOAA AR 12253. The spectral fitting is performed using the OSPEX tool \citep{Schwartz2002} within the SolarSoft (SSW) package. Though some studies suggested that microflare plasma may be multi-thermal \citep[e.g.][]{Inglis2014}, we use an isothermal model for the thermal component, given that our events are compact flares. Additionally, since compact flares are generally associated with low-altitude reconnection, a non-thermal thick-target model is also included. Each spectrum is thus fitted with a combination of these two models. Background subtraction is performed using several quiet-time intervals adjacent to the flare, with different functions applied to fit the background for different energy bands. All fittings are conducted within a time interval around the flare peak and with a single power-law model. Some key fitted parameters and corresponding uncertainties are shown in the figure, including the emission measure (EM), temperature (T) and electron spectral index ($\delta$).

In our sample, a total of 16 events have good HXR observations from either RHESSI or Fermi. Among them, 11 are identified as WLFs. Through spectral fitting, we obtain the HXR fluxes at 10 keV ($F_{10keV}$), 20 keV ($F_{20keV}$), 30 keV ($F_{30keV}$) and 40 keV ($F_{40keV}$), as well as the HXR spectral indexes ($\delta$). All these values are listed in Table  \ref{tab1}. Since the analyzed events are all very small flares, the HXR flux at a level of $10^{-2}$ $counts \cdot s^{-1} cm^{-2} keV^{-1}$ for most flares corresponds to energies generally below 30 keV (see Table  \ref{tab1}). This makes it difficult to reliably constrain the low-energy cutoff \citep[e.g.][]{Holman2011, Inglis2014}. Given this limitation, we do not attempt a quantitative assessment of the nonthermal energy for these flares.

Figure \ref{fig8} shows the scatter plots of the WL enhancements and the peak GOES 1-8 \AA\ fluxes, flare durations, HXR spectral indexes ($\delta$), and the HXR fluxes at 10 keV ($F_{10keV}$), 20 keV ($F_{20keV}$) and 30 keV ($F_{30keV}$) for all these compact WLFs. Different colors represent different types of compact WLFs. It appears that the WL enhancement is positively correlated with the peak GOES 1-8 \AA\ fluxes for type-III compact WLFs, with a correlation coefficient of $\sim$0.84, while no such correlation is found for type-I compact WLFs (panel (a)). For flare duration, the WL enhancement of type-III compact WLFs also shows a positive correlation, with a correlation coefficient of $\sim$0.81. In contrast, type-I compact WLFs exhibit a negative correlation, with a correlation coefficient of about $-0.61$ (panel (b)). This indicates that, for type-III compact WLFs, stronger energy release and longer heating duration lead to stronger WL emission. For type-I compact WLFs, however, higher energy release efficiency is required, manifested as shorter flare duration corresponding to stronger WL emission. This difference may be related to their distinct magnetic field configurations, which may differentially modulate the injection of flare energy into the lower atmosphere.

Panel (c) shows the relationship between WL enhancements and HXR spectral indices. There seems to be a positive correlation for type-III compact WLFs, with a correlation coefficient of $\sim$0.61. No correlation is seen for type-I compact WLFs. However, it should be noted that only five events are available for each type of compact WLFs, so these results require verification with a larger sample.

Panels (d)-(f) show the scatter plots between WL enhancements and HXR fluxes at 10 keV ($F_{10keV}$), 20 keV ($F_{20keV}$) and 30 keV ($F_{30keV}$). For type-I compact WLFs, the WL enhancements show a positive correlation with HXR fluxes, with correlation coefficients of $\sim$0.81, $\sim$0.78 and $\sim$0.67 at the three energies, respectively, indicating that higher HXR flux leads to stronger WL enhancement. For type-III compact WLFs, however, no such correlation is evident. It is also interesting to see that, under the same HXR flux, type-I compact WLFs appear to produce stronger WL emission than type-III compact WLFs. This difference likely stems from their distinct magnetic topologies. Type-I compact flares have a U-shaped magnetic structure, with reconnection occurring above the PIL (inside the U-shape), leading to concentrated energy release and higher heating efficiency in the lower atmosphere. In contrast, reconnection in type-III compact flares typically occurs at a null point, and the released energy is transported downward, heating the lower atmosphere at multiple locations, resulting in lower heating efficiency. This further highlights the crucial role of magnetic topology in WLF generation. It should be emphasized again that only a few events are available for both type-I and type-III compact WLFs. Therefore, these results need to be verified with a larger sample in the future.

\section{Summary} \label{sec:sum}

Currently, researches on WLFs primarily focuse on high-energy events such as M- and X-class flares. Although some C-class WLFs have been reported, most are limited to flares above C5.0. In contrast, small-scale WLFs below C5.0 remain poorly studied, with only a few documented cases, such as those by \citet{Hudson2006}, \citet{Jess2008} and \citet{Song2020}. Notably, these cases all exhibited typical compact brightening features, implying a potential intrinsic connection between compact flares and WL emission. After all such flares are characterized by low-height magnetic reconnection and energy deposition in a small-scale concentrated region, which may be physically tended to heat lower atmosphere and produce WL emission. This motivates us to conduct, for the first time, a statistical study of WL emission in compact flares.

We identified 28 compact flares including 19 C-class and 9 B-class flares. We find that 17 of the 19 C-class compact flares are WLFs, while none of the B-class flares show detectable WL enhancement. This may suggest that even in compact flares, a sufficient energy threshold is required to produce WL emission. The overall occurrence rate of WLFs in compact flares is $\sim60.7\%$ (17/28). Remarkably, this rate increases to $\sim89.5\%$ (17/19) when only considering C-class compact flares. These results strongly support our hypothesis that compact flares, particularly for those C-class and above, are easily to be WLFs.

More interestingly, we find that these compact flares can be classified into three types (Figures \ref{fig2}, \ref{fig3}, \ref{fig4}), corresponding to three different magnetic configurations (Figure \ref{fig5}), i.e., the U-shape loop (type I), the small-scale flux emergence near a sunspot (type II), and the mini fan-spine like structure (type III). We find type-I and type-III compact flares are more likely to produce WL emission with the rate of $\sim77.8\%$ (7C/(7C+2B)) and 90\% (9C/(9C+1B)), respectively. For the C-class type-I and type-III compact flares, the occurrence rates of WLFs are both 100\%. However, for the type-II compact flares (3C+6B), only one C-class flare is WLF. The occurrence rate is just $\sim11.1\%$. This may be related to the magnetic field configuration of Type-II compact flares. Under this magnetic topology, the generation of compact flare requires small-scale magnetic flux emergence and its subsequent reconnection with the background field, implying an extremely limited magnetic energy release. That may be why type-II compact flares contain the largest number of B-class flares.

It should be noted that due to the extremely small scale of these compact flares and their low reconnection height, it is difficult to obtain the accurate three-dimensional magnetic structures through the commonly used force-free field extrapolation methods. Therefore, we do not adopt the extrapolation method to determine the corresponding magnetic structures. Instead, we infer these magnetic structures by comparing the photospheric magnetic field characteristics and flare morphologies with the results of previous studies. Given the distinct differences in the observational features among the three types of compact flares, we consider the classification to be fundamentally reliable.

We examine the relationships of WL enhancements with peak GOES 1-8 \AA\ fluxes and flare durations for these compact WLFs. It is found that the WL enhancements of type-III compact WLFs positively correlate with the peak GOES 1-8 \AA\ fluxes and flare durations, indicating that stronger energy release and longer heating duration produce stronger WL emission. For type-I compact WLFs, however, there is no correlation with the peak GOES 1-8 \AA\ fluxes, but a negative correlation with flare durations, indicating that higher energy release efficiency corresponds to larger WL enhancement. This difference may arise from different magnetic field structures, which play different modulating roles in the transport and heating of flare energy to the lower atmosphere.

We also explore the relationship between HXR and WL emissions in these compact flares. The HXR emissions in these flares are weak, and most are predominated by the fluxes of below 30 keV (Table \ref{tab1}). The peak times of WL emissions in these compact flares obviously lag the peak times of HXR and SXR about $\sim1$ minutes and $\sim30$ seconds, respectively (see Figures \ref{fig2}, \ref{fig3} and \ref{fig4}). We further compare the WL emissions and the HXR spectral indices and the HXR fluxes at the energies of 10, 20, and 30 keV for these compact flares. We do not find a negative correlation between the WL enhancements and HXR spectral indices, as reported by \citet{Huang2016}. In contrast, for Type-III compact WLFs, the WL enhancement is even positively correlated with the HXR spectral index (Figure \ref{fig8}(c)). This indicates that strong WL emission can be produced in compact flares even without a large fraction of high energy electrons. However, considering only several cases of compact WLFs in our study, such results require verification with a larger sample in the future.

We also find that greater WL enhancements are associated with larger HXR fluxes for type-I compact WLFs, while no such correlation is found for type-III compact WLFs. Moreover, under the same HXR flux level, type-I compact WLFs appear to produce stronger WL emission than type-III compact WLFs (Figure \ref{fig8}(d),(e) and (f)). This may be related to their respective magnetic topologies, under which the energy transport efficiency of non-thermal electrons and their heating efficiency of the lower atmosphere differ. This further highlights the importance of the coupling between high-energy electrons and magnetic topology in the generation of WLFs.

\begin{acknowledgments}
The author sincerely thanks the referee for the valuable comments and suggestions, which helped improve the paper. This work is supported by National Natural Science Foundation of China (Grant No. 12173049, 12473053, 11803002, 11973056), National Key R\&D Program of China (Grant No. 2022YFF0503001, 2021YFA0718601) and Beijing Natural Science Foundation (Grant No. 1222029). The author thanks the SDO, GOES, RHESSI and Fermi teams for providing the data. SDO is a space mission in the Living With a Star Program of NASA.
\end{acknowledgments}


\bibliography{bibliography}{}
\bibliographystyle{aasjournalv7}

\end{document}